\documentstyle[twoside,fleqn,epsfig,latexsym,amssymb,espcrc2,graphicx,graphics]{article}

\def\bc{\begin{center}}
\def\ec{\end{center}}
\def\bs{\begin{slide}}
\def\es{\end{slide}}

\newcommand{\gev}{\mathrm{GeV}}

\newcommand{\ba}{\begin{array}}
\newcommand{\ea}{\end{array}}
\newcommand{\bes}{\begin{equation*}}
\newcommand{\ees}{\end{equation*}}
\newcommand{\beqns}{\begin{eqnarray*}}
\newcommand{\eeqns}{\end{eqnarray*}}

\def\vdir{v\kern-8.75pt\raise0.15ex\hbox{${\scriptstyle /}$}}
\def\pdir{p\kern-7.8pt\raise0.2ex\hbox{\Big{/}}}
\def\ddir{D\kern-13.75pt\raise0.15ex\hbox{\Big{/}}}
\def\partdir{\partial\kern-7.6pt\raise0.25ex\hbox{/}}
\def\ddirp{D_{\kern-5pt\perp}\kern-22pt\raise0.15ex\hbox{\Big{/}}\kern+5.5pt}

\def\ampl{{\mathcal{M}}}

\def\epe{\epsilon'/\epsilon}

\def\slash#1{\setbox0=\hbox{$#1$}\dimen0=\wd0 \setbox1=\hbox{/} \dimen1=\wd1 \ifdim\dimen0>\dimen1 \rlap{\hbox to \dimen0{\hfil/\hfil}} #1 \else \rlap{\hbox to \dimen1{\hfil$#1$\hfil}} / \fi}                                         

\def\gev{\mbox{ GeV}}

\def\op{O}

\def\lrvec#1{\setbox0=\hbox{$#1$}
    \setbox1=\hbox{$\scriptstyle\leftrightarrow$}
    #1\kern-\wd0\smash{\raise\ht0\hbox{$\raise1pt\hbox{
$\scriptstyle\leftrightarrow$}$}}\kern-\wd1\kern\wd0}

\def\spose#1{\hbox to 0pt{#1\hss}}
\def\ltapprox{\mathrel{\spose{\lower 3pt\hbox{$\mathchar"218$}}
 \raise 2.0pt\hbox{$\mathchar"13C$}}}
\def\gtapprox{\mathrel{\spose{\lower 3pt\hbox{$\mathchar"218$}}
 \raise 2.0pt\hbox{$\mathchar"13E$}}}
\def\inapprox{\mathrel{\spose{\lower 3pt\hbox{$\mathchar"218$}}
 \raise 2.0pt\hbox{$\mathchar"232$}}}

\newcommand{\nn}{\nonumber}

\newcommand{\<}{\langle}
\renewcommand{\>}{\rangle}

\newcommand{\MSbar}{\overline{\mbox{MS}}}
\newcommand{\be}{\begin{equation}}
\newcommand{\ee}{\end{equation}}
\newcommand{\bi}{\begin{itemize}}
\newcommand{\ei}{\end{itemize}}
\newcommand{\bea}{\begin{eqnarray}}
\newcommand{\eea}{\end{eqnarray}}

\newcommand{\rar}{\rightarrow}

\newcommand{\beq}{\begin{equation}}
\newcommand{\eeq}{\end{equation}}
\newcommand{\beqn}{\begin{eqnarray}}
\newcommand{\eeqn}{\end{eqnarray}}

\title{\vspace*{-0.0cm}Matrix elements of $\Delta I=3/2$ $K\rightarrow\pi\pi$ decays\thanks{Presented by Mauro Papinutto at Lattice 2003}}

\author{$\mathrm{SPQ_{CD}R}$ Collaboration:
~Ph.~Boucaud\address{\vspace*{-0.29cm}Laboratoire de Physique Th\'eorique (b\^at.210),
Universit\'e de Paris XI, 91405 Orsay Cedex, France}, 
V.~Gim\'enez\address{\vspace*{-0.29cm}Dep.~de F\'{\i}s.Te\`orica and IFIC, Univ.~de Val\`encia, Dr.~Moliner 50, E-46100, Burjassot, Val\`encia,
Spain}, 
C.-J.~D.~Lin\address{\vspace*{-0.29cm}Dep. of Phys. \& Institute for Nuclear Theory,
  University of Washington, Seattle, WA 98195-1550, USA},
V.~Lubicz\address{\vspace*{-0.29cm}Dip. di Fisica, Univ. di Roma Tre and INFN - Roma III, 
Via della V. Navale 84, I-00146 Rome, Italy},   
G.~Martinelli\address{\vspace*{-0.29cm}Dip. di Fisica, 
Universit\`a ``La Sapienza" and INFN-Roma I, P.le A. Moro 2, I-00185 Rome,
Italy},
M.~Papinutto\address{\vspace*{-0.29cm}DESY-Hamburg, Theory Group, Notkestrasse
 85, D-22607 Hamburg, Germany},
C.~T.~Sachrajda\address{\vspace*{-0.0cm}Department of Physics and Astronomy, University of
Southampton, Southampton SO17 1BJ, England}}
\begin{document}

\begin{abstract}
\vspace*{-0.2cm}
We present  a numerical computation of matrix elements of $\Delta I=3/2$
$K\rar \pi \pi$ decays by using Wilson fermions. In order to extrapolate 
to the physical point we work at unphysical kinematics and we resort to
Chiral Perturbation Theory at the next-to-leading order. In particular we
explain the case of the electroweak penguins $\op_{7,8}$ which can
contribute significantly in the theoretical prediction of $\epe$.
The study is done at $\beta=6.0$ on a $24^3\times64$~lattice.

\vspace*{-0.4cm}
\end{abstract}
 
\maketitle

\section{INTRODUCTION}
\vspace*{-0.1cm}
The study of non-leptonic kaon decays is a challenging but important task,
which is underlined by the recent accurate measurement of a non-zero
$\epsilon^{\prime}/\epsilon$ parameter \cite{epsilonp} (the first
confirmed observation of direct CP violation) and the long-standing 
puzzle of the $\Delta I = 1/2$ rule. 

For example, the electroweak penguins (EWPs) 

\vspace*{-0.4cm}
\bea
 \op_{7} &=& \frac{3}{2}(\bar{s}_{\alpha}d_{\alpha})_{L} 
       \sum_{q} e_{q} (\bar{q}_{\beta} q_{\beta})_{R},\nn \\ 
 \op_{8} &=& \frac{3}{2}(\bar{s}_{\alpha}d_{\beta})_{L}
    \sum_{q} e_{q} (\bar{q}_{\beta} q_{\alpha})_{R} ,\nn
\eea

\vspace*{-0.3cm}
\noindent although suppressed by the factor $\alpha_{\mathrm{em}}/\alpha_{s}$ 
when compared to the QCD penguins, 
are enhanced by the $\Delta I=1/2$ rule and tends to cancel 
significantly the contribution of the QCD penguins 
in the theoretical prediction of $\epsilon^{\prime}/\epsilon$.

Here we present the first lattice
determination of the $\Delta I=3/2$ $K\rar \pi \pi$ matrix elements (MEs)
of the EWPs obtained directly with a two-pion final state. The strategy,
explained in~\cite{lattice0,lattice}, consists in
computing $\ampl^{\tiny\textrm{(7,8)}}_{\tiny\textrm{kin}}\equiv_{I=2}\<\pi\pi|\op_{7,8}|K^0\>$ at some unphysical kinematics
on the lattice (called SPQR kinematics) and then using next-to-leading 
order (NLO) Chiral Perturbation
Theory ($\chi$PT) to obtain the MEs at the physical point.

\section{OPERATOR MATCHING}

\vspace*{-0.1cm}
We now discuss the problem of the matching of lattice-regularised
operators to the continuum renormalization scheme in
which the Wilson coefficients are calculated. 
Even if chirality is preserved by the regularization, $\op^{3/2}_{7}$
and $\op^{3/2}_{8}$
(the $\Delta I=3/2$ parts of $\op_{7,8}$) mix between themselves. 
In a Wilson-like regularization there is in general also mixing 
with operators of different naive chiralities. In the 
$\Delta I=3/2$ case, $SU(2)$ isospin symmetry forbids mixing with 
lower dimensional operators while $\cal CPS$ symmetry implies that mixing with
other dimension six operators with different chiral properties occurs
only in the parity conserving sector~\cite{rimom}. The mixing is thus:

\vspace*{-0.55cm}\bea
 \bigg ( \begin{array}{cc} \op^{3/2}_{7}(\mu)\\
    \op^{3/2}_{8}(\mu) \end{array} \bigg ) =
 \hat{{\cal Z}}(\mu a)
 \bigg ( \begin{array}{cc} \bar{\op}^{3/2}_{7}(a)\\
                        \bar{\op}^{3/2}_{8}(a) \end{array} \bigg )
\eea

\vspace*{-0.14cm}\noindent where $\mu$ is the renormalization scale. 
With an obvious
notation we define ${\cal Z}_{77}(\mu a)$, ${\cal Z}_{78}(\mu a)$,
${\cal Z}_{87}(\mu a)$, ${\cal Z}_{88}(\mu a)$ to be the MEs of 
$\hat{{\cal Z}}(\mu a)$.

\begin{figure*}[htb]
\vspace*{-1.5cm}\hspace*{-0.5cm}\mbox{\epsfig{figure=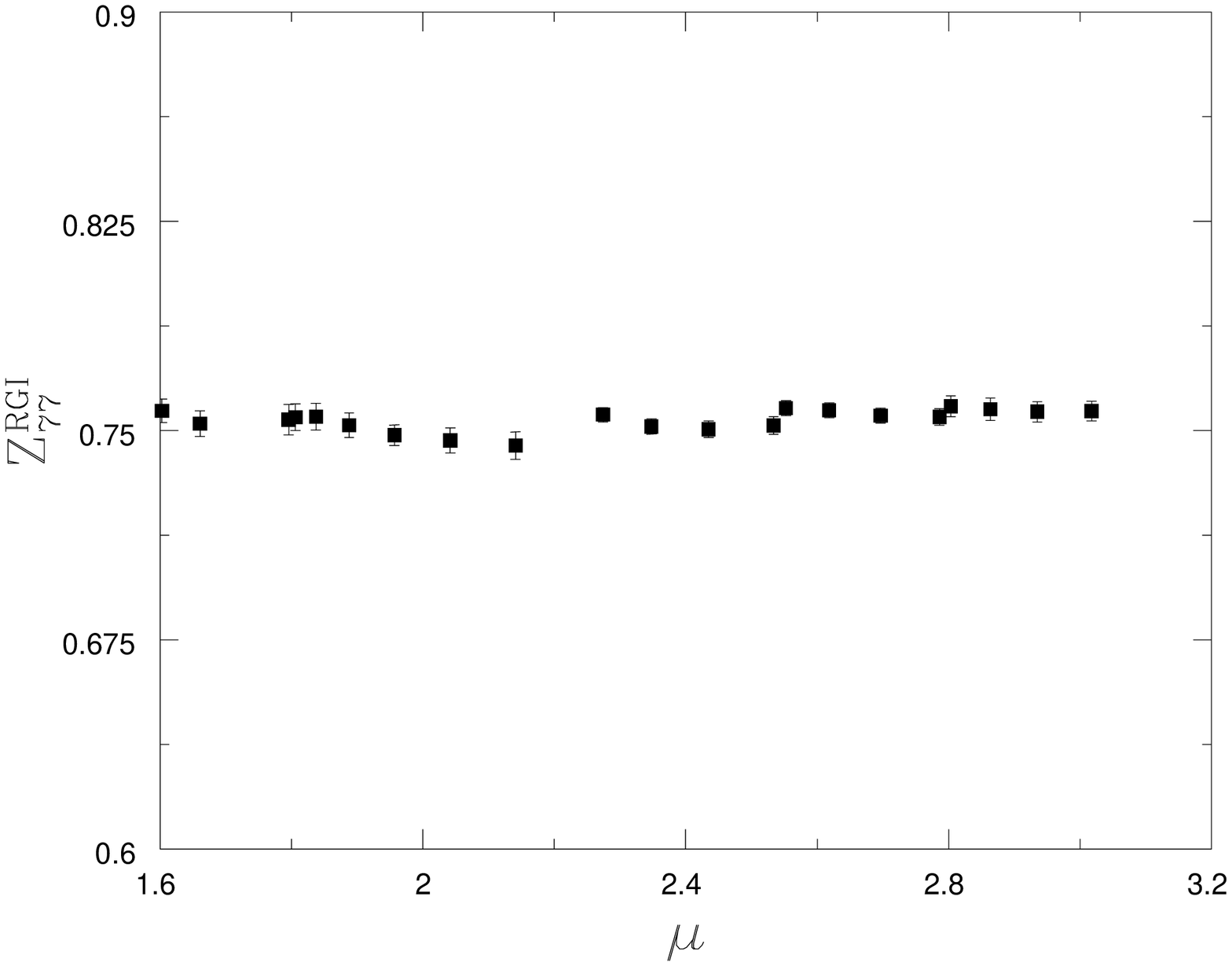,angle=0,width=0.35\linewidth}\put(-5,0){\epsfig{figure=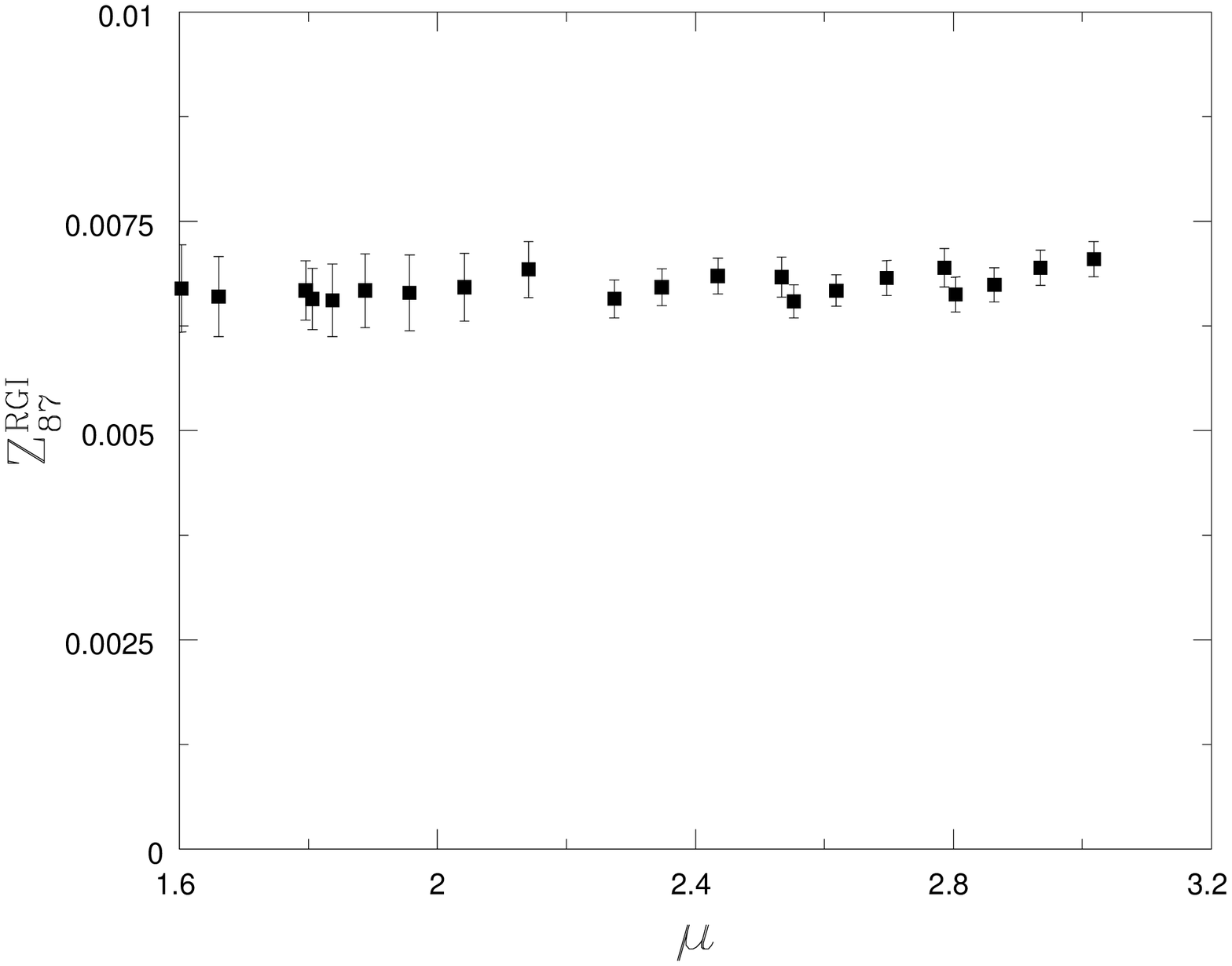,angle=0,width=0.35\linewidth}}\put(150,0){\epsfig{figure=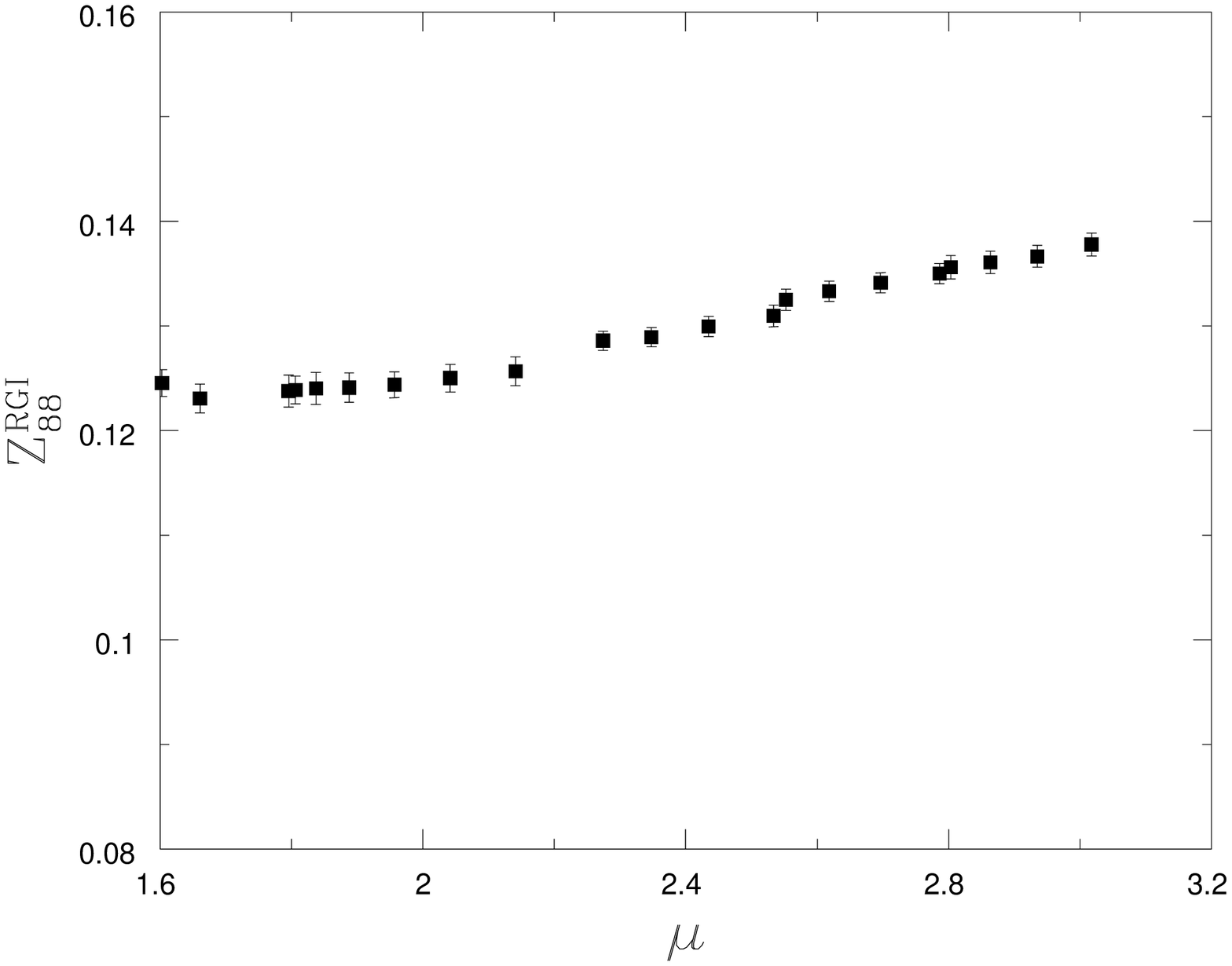,angle=0,width=0.35\linewidth}}}
\vspace*{-1.0cm}
\caption{${\cal Z}^{RGI}_{77}$, ${\cal Z}^{RGI}_{87}$, ${\cal Z}^{RGI}_{88}$
from the non-perturbatively determined $\hat{{\cal Z}}(\mu)$ as defined in Eq.~\ref{eq:RGI}.}
\label{fig:RGI}
\vspace*{-0.48cm}
\end{figure*}

To compute $\hat{{\cal Z}}(\mu a)$ we use the (mass independent) 
RI-MOM renormalization scheme which can be easily implemented 
in a non-perturbative way, also in the case of four fermion 
operators~\cite{rimom}. Spurious effects due to the Goldstone boson
contamination have been removed by using the method explained 
in~\cite{bbbar,lattice0}.

Once computed $\hat{{\cal Z}}(\mu a)$ non-perturbatively, 
we have to check to what extent it follows the renormalization 
group behaviour predicted by perturbation theory at the NLO. We thus compute  
\bea
\label{eq:RGI}
\hat{\cal Z}^{RGI}(a)& =& (\hat{w}^{-1})^T(\mu)\hat{\cal
Z}(\mu)\quad\textrm{with}\nn\\(\hat{w}^{-1})^T(\mu)&=& 
\left[\alpha_s (\mu) \right]^{- \frac{\hat \gamma^{\scriptsize{(0)}}}
{2\beta_{0}}}\left[1+{\alpha_s(\mu)\over 4\pi} \hat J^T\right]
\eea

\vspace*{-0.2cm}\noindent
where $\hat{J}$ has been computed in~\cite{scimemi}. 
$\hat{\cal Z}^{RGI}$ should not depend on the scale $\mu$ (see Fig.~\ref{fig:RGI}).

The observation that no clear $O(a^2p^2)$ 
lattice artifacts are visible in ${\cal Z}_{77}$, ${\cal Z}_{87}$ 
together with the very large anomalous
dimension of ${\cal Z}_{78}$, ${\cal Z}_{88}$ (6 and 16
at LO) suggest that higher orders in PT could contribute significantly in
particular to the RG evolution of ${\cal Z}_{88}$.

The central value for the renormalization constants at a given scale
$\mu$ is extracted in the following way: we
compute $\hat{\cal Z}^{RGI}$ by fitting the plateaux to a
constant in the interval $2.2\gev\leq \mu\leq 2.6\gev$. We then run 
$\hat{\cal Z}^{RGI}$ down
to the desired scale $\mu$. The systematic error is estimated
by computing the variation of the value obtained at this scale 
from the evolution of the original values (those used to build the
combination of Eq.~\ref{eq:RGI} and Fig.~\ref{fig:RGI}) at 
the different scales~$\mu_0$. 

\vspace*{-0.1cm}
\section{USE OF THE CHIRAL EXPANSION}

\begin{figure*}[htb]
\vspace*{-0.3cm}\hspace*{-0.2cm}\mbox{\epsfig{figure=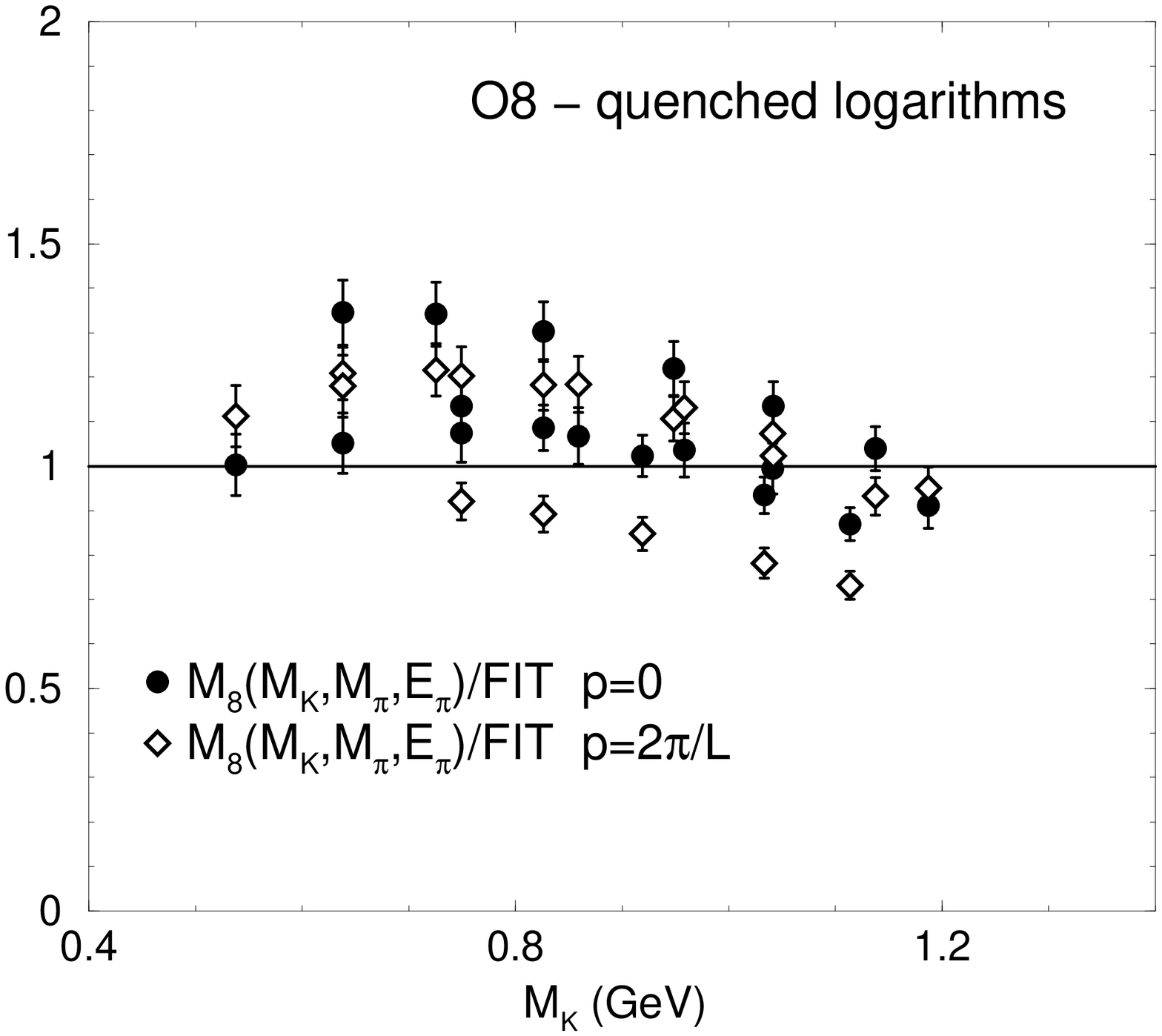,angle=-90,width=0.33\linewidth}\put(7,+1){\epsfig{figure=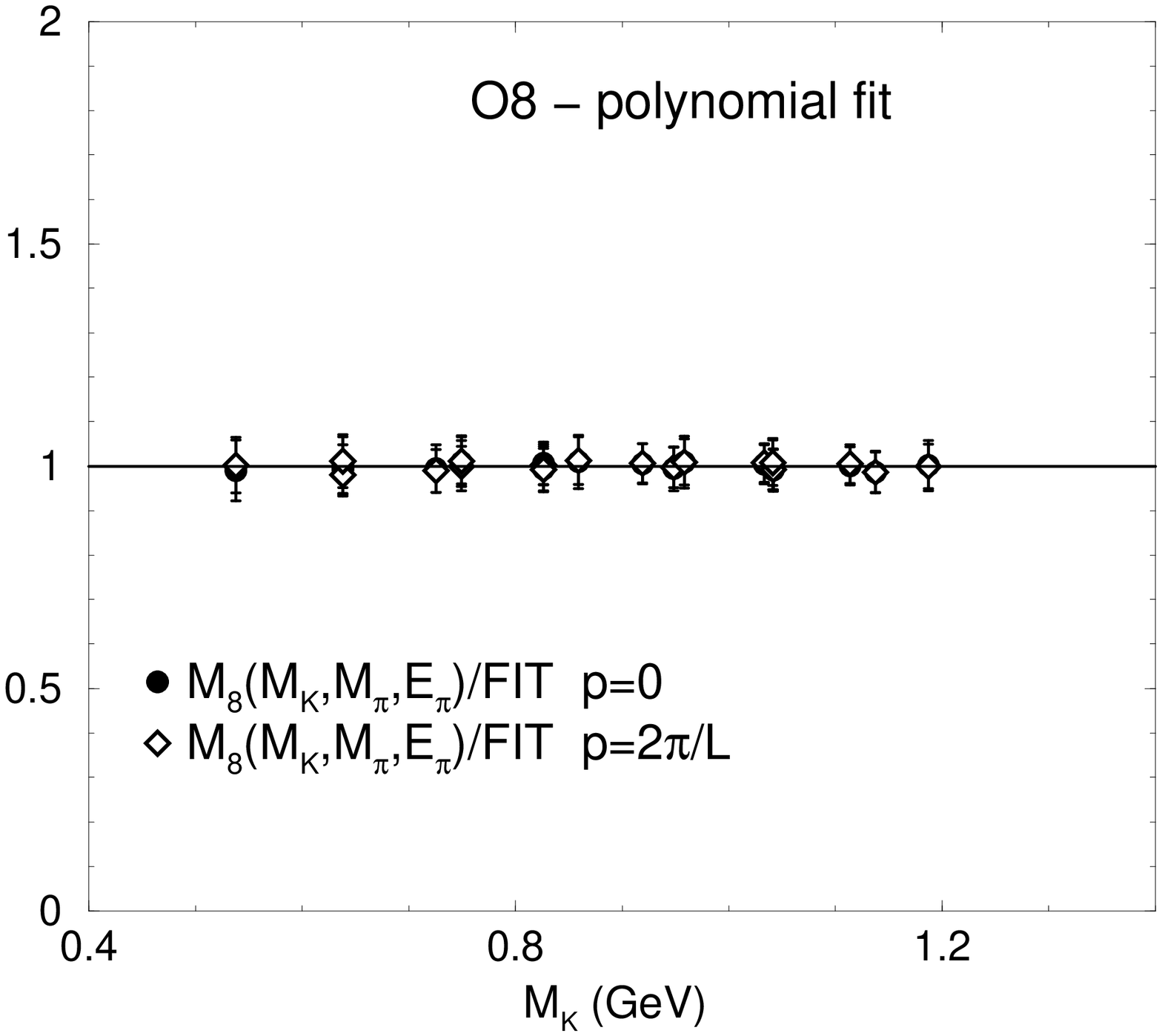,angle=-90,width=0.33\linewidth}}\put(164,0){\epsfig{figure=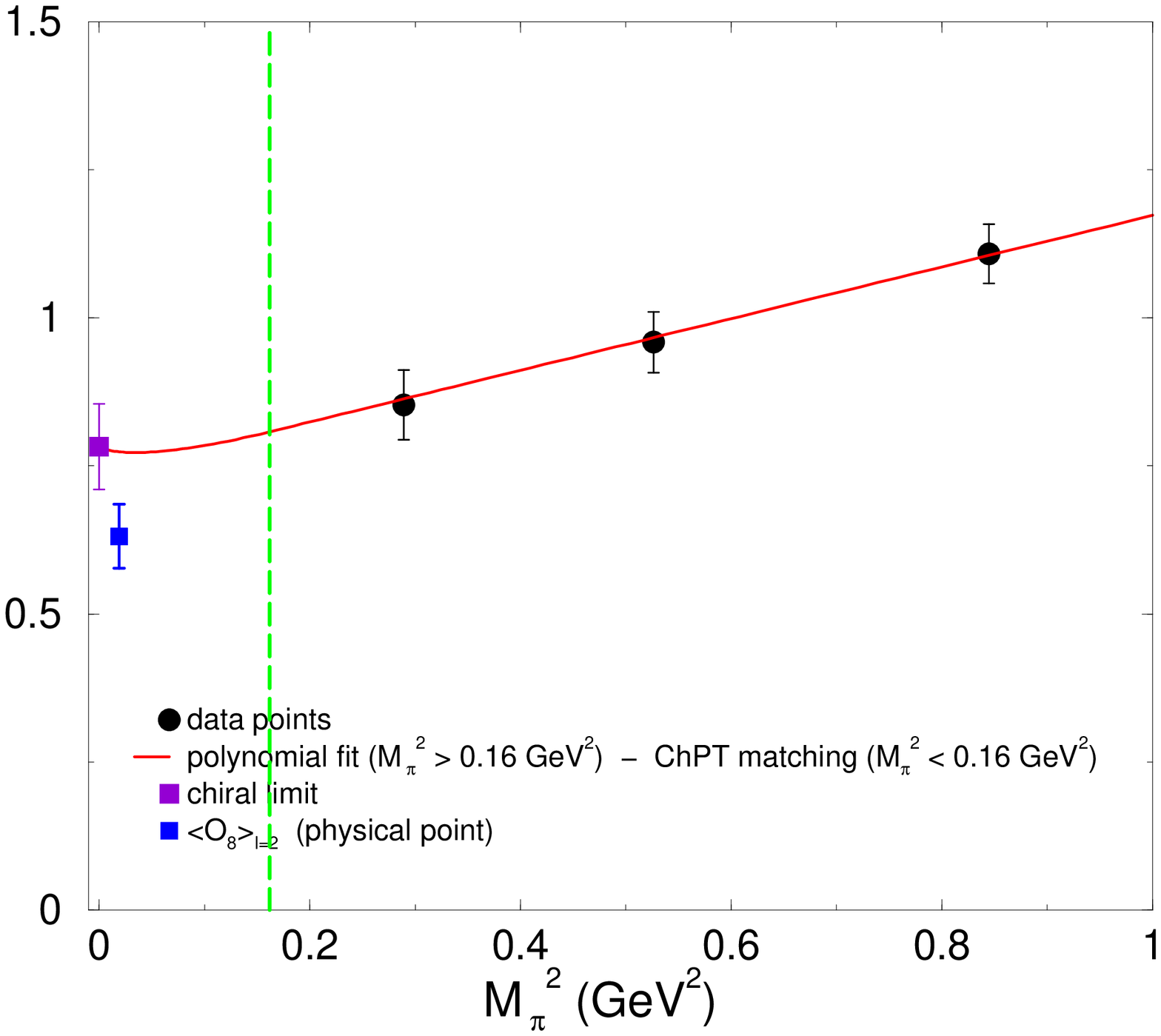,angle=-90,width=0.35\linewidth}}}
\vspace*{-1.0cm}
\caption{(a)-(b) Quality of the q$\chi$PT and polynomial fits for
  $\op_8$. (c) $\chi$PT-polynomial matching for $\op_8$ at
  $M_\pi=E_\pi=0.4\gev$, $M_K=0.41\gev$ (only data with $M_\pi=M_K$ are plotted).} 
\label{fig:fits}
\vspace*{-0.48cm}
\end{figure*}


\begin{table}[h]
\caption{Results in $\gev^3$ for the MEs at the physical point (the
  operators are renormalized in $\MSbar$-NDR at $\mu=2\gev$)  
obtained by using only points with $M_K,E_\pi<\Lambda$. 
In the quenched case the values $\alpha=0.1$ and $m_0=0.5\gev$ have been used.}
\label{tab:fits}
\hspace*{-0.0cm}{\footnotesize\begin{tabular}{cc|cc|cc}\hline\hline
fit & $\Lambda$ & $\langle O_7\rangle_{\tiny\textrm{$I\!\!=\!\!2$}}^{\tiny\textrm{phys}}$ & $\frac{\chi}{\tiny\textrm{d.o.f}}$ & $\langle O_8\rangle_{\tiny\textrm{$I\!\!=\!\!2$}}^{\tiny\textrm{phys}}$ & $\frac{\chi}{\tiny\textrm{d.o.f}}$ \\
\hline
\hline
poly. &$1.2$ &  0.120(11) & 0.023&0.714(62) & 0.031\\ 
poly. &$0.8$ &  0.125(13)& 0.012 &0.715(75) & 0.037\\ 
\hline
$\chi$PT &$1.2$ & 0.143(13)& 0.817   & 0.836(73)& 0.815\\ 
$\chi$PT &$0.8$ & 0.123(13)& 0.482   & 0.700(72)& 0.626\\ 
\hline
q$\chi$PT &$1.2$ & 0.254(25)&  13.2    & 1.49(16) &  10.8\\   
q$\chi$PT &$0.8$ & 0.212(24)&  5.95    & 1.19(14) &  4.91\\   
\hline 
\hline
\end{tabular}}
\vspace*{-0.6cm}
\end{table}

\vspace*{-0.1cm}
The NLO contribution in $\chi$PT, both
in the quenched approximation and in the full-QCD case, has been
computed for the SPQR kinematics in~\cite{lattice1}. 
One can thus think of extracting the low energy
constants (LECs) of the LO and NLO (in the case of the EWPs 
${\cal O}(p^0)$ and ${\cal O}(p^2)$ respectively) from lattice data and to use
them in the NLO formula for the physical kinematics. Both forms can
be derived from the formula in the most general kinematics 
(energy momentum injection via the weak operator allowed but with all
the particles on-shell) which reads

\vspace*{-0.6cm}
\bea
(\ampl^{\tiny\textrm{(7,8)}}_{\tiny\textrm{gen}})^{\tiny\textrm{$\chi$PT}}
 = A^{\chi} (1 + [\chi\;
  {\mathrm{logs}}]_{\tiny\textrm{gen}}) + B_1^{\chi} M^{2}_{K}
 +\nn\\B_3^{\chi} M^{2}_{\pi} + B_2^{\chi} 
   (p_{\pi^{+}}+p_{\pi^{0}})\cdot p_{K}
 + B_4^{\chi} (p_{\pi^{+}}\cdot p_{\pi^{0}}) 
\label{eq:ChPT}
\eea

\vspace*{-0.2cm}\noindent 
where we have symmetrized over $p_{\pi^{+}}$,$p_{\pi^{0}}$ since the two
pion state is a pure $I=2$ state.

The SPQR kinematics corresponds to put the $K$ and one $\pi$ at rest
while the other $\pi$ is either at rest or has the minimum momentum
allowed on the lattice. The previous formula can thus be re-expressed in
terms of $M_K$, $M_\pi$ and $E_\pi$, where $E_\pi$ is the energy
of the (possibly) moving pion. 

Besides the fit with the complete form predicted by $\chi$PT 
(Eq.~\ref{eq:ChPT}),
we also try a fit with the polynomial part only (the coefficients of
which are now ``effective'' LECs) which we call
$(\ampl^{\tiny\textrm{(7,8)}}_{\tiny\textrm{gen}})^{\tiny\textrm{poly}}$.
Tab.~\ref{tab:fits} reports the results of the fits. 
It is clear by looking at the $\frac{\chi^2}{\tiny\textrm{d.o.f}}$ 
that our data are not in the kinematical range where the chiral 
logarithms are visible. This is 
visualized in Fig.~\ref{fig:fits}.a-b.
A possible explanation for this behaviour is that, since logarithms come from
loops of Goldstone bosons (GBs), they become relevant at a 
scale of the order of the mass of the GBs. Since the contribution
of the counterterms is instead controlled by the physics at the scale of the
cut-off (i.e. ${\cal O}(1\gev)$), it is dominant in the kinematical 
region accessible on the lattice (in our simulation  
$M_\pi, M_K, E_\pi \in [0.54,1.2]\gev$).

\begin{table}[b]
\vspace*{-0.7cm}
\hspace*{-0.3cm}{\footnotesize\begin{tabular}{c|cc|cc}\hline\hline
 $M_\pi,M_K$ & \multicolumn{2}{c|}{$\langle O_8\rangle_{\tiny\textrm{$I\!\!=\!\!2$}}^{\tiny\textrm{phys}}$}
& \multicolumn{2}{c}{$\langle O_7\rangle_{\tiny\textrm{$I\!\!=\!\!2$}}^{\tiny\textrm{phys}}$}\\
\hline
\hline
poly. & \multicolumn{2}{c|}{$0.714(62)$}& \multicolumn{2}{c}{$0.120(11)$}\\
\hline
\hline
match. & $\chi$PT & q$\chi$PT & $\chi$PT & q$\chi$PT\\ \hline
0.3,0.5 & $0.697(60)$ & $0.738(64)$&$0.117(11)$& $0.125(12)$\\
0.4,0.5 & $0.664(57)$ & $0.758(66)$&$0.111(10)$& $0.128(12)$\\
0.3,0.31& $0.638(55)$ & $0.711(61)$&$0.106(10)$& $0.120(11)$\\
0.5,0.51& $0.635(55)$ & $0.825(72)$&$0.106(10)$& $0.141(13)$\\
\hline 
\hline
\end{tabular}}
\caption{Results (same units and scheme of Tab.~\ref{tab:fits}) for the
  matching procedure at the point ($M_\pi=E_\pi$,$M_K$).}
\label{tab:match}
\vspace*{-0.0cm}
\end{table}

On the other hand, one expects that predictions of $\chi$PT (including
the logs.) are valid when all the energy scales are sufficiently
small (let's say below the mass of the Kaon). If we assume that 
there exists a region in which the two
descriptions of the form factors are reasonably close to each other, 
we can perform the matching between these two forms~\cite{ryan}. This
amounts to fit the data with $(\ampl^{\tiny\textrm{(7,8)}}_{\tiny\textrm{gen}})^{\tiny\textrm{poly}}$ and then to impose the equality of  
$(\ampl^{\tiny\textrm{(7,8)}}_{\tiny\textrm{gen}})^{\tiny\textrm{poly}}$ with
$(\ampl^{\tiny\textrm{(7,8)}}_{\tiny\textrm{gen}})^{\tiny\textrm{$\chi$PT}}$
 and of their
first derivatives with respect to the kinematical variables 
$M^{2}_{K}$, $M^{2}_{\pi}$, $p_{\pi^{+}}\!\!\cdot\! p_{K}$, $p_{\pi^{+}}\!\!\cdot\!
p_{\pi^{0}}$ 
at a matching point (chosen in the SPQR kinematics).

Results for different matching points are shown in
Tab.~\ref{tab:match}. Lattice data have not yet been corrected for all
of the finite volume corrections explained in~\cite{lattice0} and so our
results are still preliminary. Nevertheless it's interesting to study the
systematics of the $\chi$PT-polynomial matching and the contribution of
the NLO at the physical point. Since quenched logarithms are very
different from the full QCD ones (and assuming that quenching 
does not change drastically the kinematical behaviour of the form 
factors in the range accessible to the lattice) we think that the most
sensible way of extrapolating to the physical point is to perform
the matching with full $\chi$PT. If we choose the matching point 
{\small($M_\pi,M_K$)=(0.4,0.5)$\!\!\gev$} we get:
{\small\bea\hspace*{-0.3cm}\langle
\op_8\rangle=0.664(57)\left(^{+40}_{-38}\right)\left(^{+50}_{-40}\right)\quad
\langle\op_7\rangle=0.111(10)\left(^{+6}_{-4}\right)
\left(^{+9}_{-6}\right)\nn
\eea}
\noindent where the first error is statistical, 
the second is the systematics of
non-perturbative renormalization and the third is the systematics due to the
variation of the matching point.
At the physical point, the NLO gives a negative contribution of order
$15\div22\%$ with respects to the LO.


\vspace*{-0.0cm}
\hspace*{-0.4cm}
\parbox{7.5cm}  
\end{document}